# А.Г. КОЛОНИН[1]*, В.Г. КРЮКОВ[2]

[1]Новосибирский Государственный Университет
[2]Домодедово Бизнес Партнер
akolonin@gmail.com

# ВЫЧИСЛИТЕЛЬНАЯ КОНЦЕПЦИЯ ПСИХИКИ

В статье приводится обзор подходов по моделированию человеческой психики в перспективе построения искусственной. На основе обзора предлагается концепция когнитивной архитектуры, где психика рассматривается в качестве операционной системы живого или искусственного субъекта, включая пространство состояний, обуславливающее его жизненные смыслы в связи со стимулами внешнего мира, и интеллект как систему принятия решений для действий в отношении этого мира с целью удовлетворения указанных потребностей. На основе концепции предлагается вычислительная формализация для создания систем искусственного интеллекта посредством обучения на основе опыта в пространстве множества потребностей, с учетом их биологической либо экзистенциальной значимости для интеллектуального агента. Тем самым, формализуется задача построения общего искусственного интеллекта как системы принятия оптимальных решений в пространстве специфических для агента потребностей в условиях неопределенности, с максимизацией успехов в достижении целей, минимизации экзистенциальных рисков и максимизации энергоэффективности. Также приводится минимальная экспериментальная реализация модели.

**Ключевые слова:** мотивация, принятие решений, пространство потребностей, пространство состояний, психика, эмоции, энергоэффективность

## Введение

Попытки оцифровать психику или построить ее системную модель, которая была бы к тому же и вычислимой, предпринимались многими учёными, начиная с Норберта Винера [1]. Но именно Винер сформулировал основную проблему в поиске решения – сложность междисциплинарных знаний и коммуникации. Как мы покажем дальше, для создания системной модели требуются знания психологии (психоанализа), системного анализа и производственной рыночной экономики (микроэкономики). А чтобы выстроить моделирующий психику программный комплекс нужно знание системного программирования.

На наш взгляд, построение искусственной психики, как операционной управляющей системы системы неразрывно связано с решением задачи создания искусственного интеллекта (ИИ), равного человеческому, или превосходящего человеческий. Потому что человеческий интеллект это управляющая система его деятельности, решающая основные задачи выживания и воспроизводства с учётом горизонта планирования. Для осуществления подобной деятельности система, будь то человек или ИИ, должна принимать оптимальные решения в каждый момент и в течении всего жизненного цикла. А оптимальность решений может быть определена только в рамках системы потребностей, обусловленных средой существования носителя самого интеллекта и его собственной физической организацией, то есть - в рамках психики, объединяющей совокупность мыслительных процессов субъекта, внутренний мир его мотиваций и отражение в этом мире окружающей реальности.

Попытка описать работу психики сделана в монографии Леонтьева Д.А. [2]. В работе проанализированы труды классического психоанализа и научные открытия советской и русской школы. Согласно ей, работа психики основана на управлении жизнедеятельностью с целью удовлетворения потребностей разных уровней в соответствии с их приоритетами в текущем моменте. "Функциональную роль психики в наиболее общем виде можно охарактеризовать как регуляцию жизнедеятельности на основе ориентировки в объективном мире посредством построения субъективных образов действительности" [2].

## Обзор подходов

В соотвествии с обобщенным определением интеллекта, данным Герцелем [3] и Вонгом [4], «интеллект это способность достигать сложных целей в сложных условиях при ограниченных ресурсах». При этом, если удовлетворение условию ограниченных ресурсов можно рассматривать как одну из подцелей во внутренней иерархии сложной составной глобальной цели, то формально можно наличие интеллекта у системы свести к возможности осуществлять много-параметрическую оптимизацию при решении задачи целенаправленного управления в некоей операционной среде. При этом уровень интеллекта будет определяться количеством и сложностью структуры пространства параметров данной операционной среды и стоящих перед системой целей.

Другими словами, задача интеллекта это находить оптимальные решения на разных уровнях взаимодействия с окружающим миром, безотносительно к тому, является ли система живой (животное или человек) или неживой (ИИ).

Согласно Канеману [5] управляющая система человеческой деятельности состоит из двух частей: медленной (осознанное мышление) и быстрой (рефлекторной или интуитивной). Сознание относится к медленной части и практически не контролирует жизнедеятельность. Сознание принимает часть решений и воспринимает лишь часть внешней и минимум внутренней информации о состоянии системы.

Принятие решений в живой системе, а также формирование опыта ею, или ее обучение, осуществляется посредством эмоций [6, 7]. "Эмоции, таким образом, сообщают нам на своем специфическом «языке» непосредственную оценку значимости сопровождаемых ими образов, то есть оценку отношения соответствующих объектов и явлений к реализации потребностей субъекта" [2].

Эмоции это реакция психики на удовлетворение потребности или достижение (либо не достижение) цели. Эмоция переводит систему в новое состояние, компенсируя отклонение от равновесного состояния. Положительная эмоция возникает в системе, когда происходит удовлетворение потребности или цели. Отрицательные эмоции возникают, когда потребность требует удовлетворения. Это становится информационным сигналом для системы о повышении приоритета потребности. Болевые сигналы выполняют ту же функцию на физиологическом уровне. Таким образом, эмоции можно использовать в качестве функции подкрепления при обучении, причем яркость эмоции может рассматриваться как значение функции подкрепления.

Для создания модели интеллекта, включая реакцию системы на внутренние мотивирующие стимулы и восприятия реакций внешнего мира, в перспективе имеющихся ожиданий и собственных воздействий на окружающую среду, предполагается применять теорию функциональных систем (ТФС) Анохина [8] и принцип динамического равновесия Берталанфи [9]. На основе иерархической модели ТФС может быть построена вычислительная концепция и реализующая ее когнитивная архитектура, описывающая поведение как автономного интеллектуального агента (ИА) [10], так и социума в целом [11].

Психика это операционная система по управлению жизнедеятельностью человека, либо гипотетической искусственной системы. Интеллект это часть системы по принятию решений, включающей подсознательную и сознательную составляющие, действующие комплементарно, на конкурентной основе [12]. Подсознание принимает рефлекторные решения (как это имеет место в случае человеческой интуиции либо с использованием современных систем на основе глубоких нейронных сетей), а мышление – стратегические осознанные решения в рамках горизонта планирования с учётом доступности ресурсов.

Процесс принятия решений для достижения текущих целей, для удовлетворения насущных потребностей физиологических и психологических, может быть описан как мотивация на совершение тех или иных действий под воздействием эмоций. Мотивация деятельности определяется как внешними воздействиями, так и внутренним состоянием системы. Внутреннее состояние определяется приоритетом физиологических и психологических потребностей, а также уровнем доступных ресурсов энергии и внешних физических ресурсов. Воздействия внешней среды, а также данные о состоянии подсистем (органов живого организма) воспринимаются как ощущения сенсорами системы и передаются как восприятия для изменения внутреннего ее состояния. Принятые, в процессе мотивации, решения передаются на исполнение актуаторам системы для воздействия на окружающую среду либо внутренние подсистемы (органы живого организма).

Ведь деятельность, как уже говорилось, связана обычно не с одной, а с целым рядом потребностей. Общий смысл конкретной деятельности — это сплав ее частичных смыслов, каждый из которых отражает ее отношение к какой-либо одной из потребностей субъекта, связанной с данной деятельностью прямо или косвенно, необходимым образом, ситуативно, ассоциативно или как-либо иначе. Поэтому деятельность, побуждаемая всецело «внешними» мотивами — столь же редкий случай, как и деятельность, в которой они полностью отсутствуют [2].

В текущий момент система выбирает наиболее эффективное решение для удовлетворения актуальных потребностей, если их приоритет высокий. Либо находит такой вариант деятельности, чтобы эффективно удовлетворить будущие потребности, которые непременно возникнут. Интеллектуальная система может знать, что организму через несколько часов понадобится пища, но точное время наступления голода будет зависеть от расхода энергии на деятельность в ближайшее время. Поэтому интеллектуальная система стремится как обеспечить необходимые ресурсы, которые скоро понадобятся, так и экономить их потребление, если это возможно.

Соображения вычислительной и тем более экономической эффективности рассматривались как ключевые при создании вычислительных архитектур ИИ и ранее [13, 14]. Однако существующие подходы в ИИ как на основе максимизации вероятностных предсказаний [15], так и на основе различных видов вероятностной логики [16, 17], в явном виде соображения эффективности не рассматривают. Вместе с тем, в последнее время также Дэвид и Ле Кун [18] указывает на необходимость введения понятия энергоэффективности как фундаментального для построения архитектур ИИ. При этом, следует отметить, что имеющиеся модели вероятностной логики допускают неявный учет факторов энергоэффективности в качестве дополнительных переменных [16, 17].

Отличие предлагаемой концепции от как чисто вероятностных моделей [16, 17], так и от моделей, где ресурс-потребление является фактором, ограничивающим логический вывод [13, 14], состоит в том, что решение предполагается направленным на конечный результат в перспективе управления рисками, как экзистенциальными (угроза существованию системы или социуму, в который она входит), так и энергетическими или экономическими, в свете теории перспектив [19].

Согласно П.К. Анохину [8], реальный интеллект ориентирован на конечный результат, о чем писал академик. В связи с этим, мы полагаем, что принятие решений может осуществляться с использованием экономических методов затратного анализа, с учетом, в том числе, вероятностей экзистенциальных угроз и возможных затрат на их избегание. Поэтому решение может выбирается по критериям наибольшей полезности, максимальной безопасности и минимальной затратности (коэффициент полезного действия для получения энергии выживания) для системы на основе мульти-целевого анализа. К этому подошел академик Анохин, но модели марксистской плановой советской экономики, использованные им, не могли отражать конкуренцию задач в многоцелевых системах. С применением принципов рыночной экономики модель интеллекта может оказаться формализуемой и вычислимой.

Общая теория систем Людвига фон Берталанфи [9] предполагает, что живая система находится в постоянном динамическом равновесии. Внешние воздействия или внутренние изменения отклоняют систему от равновесного состояния. Любое отклонение требует возврата назад в равновесное состояние. Для этого требуется совершить действие, тратя энергию системы. А энергию надо восполнить. Решения такой задачи, как мы полагаем можно найти в экономических многоцелевых моделях управления рисками [19].

Для обеспечения выживания вида, социума или индивида человек осуществляет целенаправленную деятельность. Деятельность требует затрат энергии, которые получают из пищи. Обеспечение пищей - одна из главнейших задач, которую решают все живые организмы, наряду с избеганием экзистенциальных угроз.

Для обеспечения пищей также требуется совершать деятельность. Таким образом, мы получаем глобальный замкнутый цикл жизнедеятельности. В основе которого базируется энергия выживания. Следовательно, для оценки целенаправленной деятельности должно использовать «энергию выживания». Но не прямом физическом смысле, а как «универсальную валюту» (что созвучно с идеей «энергорубля» или «токена» цифровой крипто-экономики) для сравнения, как разноплановую условную единицу, которая дает количественную оценку для физиологических и психологических процессов вычислительной модели живого организма, либо искусственной психики как операционной системы для системы ИИ.

Энергия выживания - это биологическая (в случае живого организма) либо электрическая (в случае искусственной системы) энергия, необходимая организму или системе для осуществления целенаправленной внешней деятельности и исполнения внутренних процессов.

Об энергии организма и психики писал Бехтерев [20]. Он описал разные типы энергии в организме и принципы их взаимодействия. В начале XX века Фрейд в книге "По ту сторону принципа удовольствия" [21] написал, что психика работает по экономическим принципам.

Внутреннее состояние системы описывается в пространстве потребностей, или «матрице потребностей» в [22], являющемся основным элементом модели живой интеллектуальной системы (например, человека) или вычислительной когнитивной архитектуры ИИ.

Обучение системы, в рамках предлагаемой концепции, осуществляется посредством приобретения опыта, для обеспечения принятия решений по удовлетворения потребностей системы в рамках горизонта планирования. В случае человека, человеческое знание состоит из огромного количества элементарного опыта, полученного из собственных действий или из текстов (осмысленный опыт других людей). Это знание, определенное в пространстве потребностей, структурируется в покрытие Маркова [23]. Самообучение это получение и осмысление опыта с выбором наиболее эффективного с точки зрения затрат и получаемой пользы.

Основа формирования опыта в психике открыта Адлером [24] в развитии личности. Тенденции по увеличению значимости (психологической ценности) целей или потребностей можно понять из его авторской идеи "комплекс неполноценности". А введённое им понятие психологической ценности позволило сравнивать и рассчитывать психологические и физиологические потребности.

### Концептуальная архитектура

В расширительной трактовке, интеллект – это система анализа и принятия решений на основании знаний и опыта для обеспечения эффективной функциональной деятельности человека, социума, либо высокоуровневой интеллектуальной системы (обладающей искусственным интеллектом) или совокупности таких систем. Причем, интеллект это не абстрактный набор закономерностей физического мира или получаемых из него фактов, а способность системы управлять деятельностью на основе осмысленного опыта. На сегодняшний день интеллект существует только в живых системах, но мы рассматриваем возможность его создания в искусственных системах.

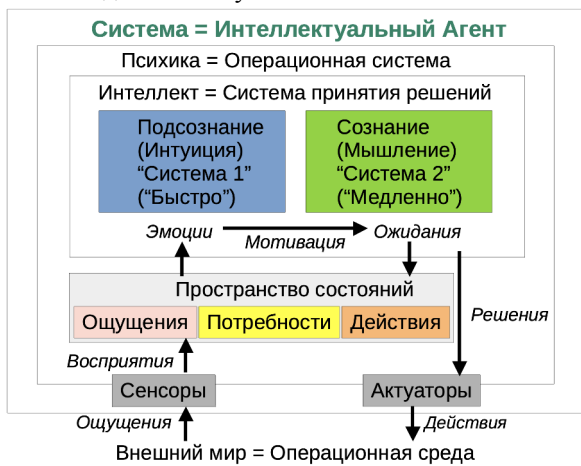

Рис. 1. Концептуальная когнитивная архитектура искусственной психики

Для этого мы предлагаем следующую антропоцентричную концептуальную архитектуру психики, включающую пространство потребностей, связывающее интеллектуальную систему с ее внешним миром, с учетом физических особенностей ее собственной реализации, с системой принятия решений в этом пространстве (Рис. 1). Данная концепция позволяет перейти к формализации и апробации математической модели, обсуждаемой ниже.

Интеллект человека развивается ежесекундно с момента рождения. Значит, возможно построить модель интеллекта также начиная с самого простого состояния. Человеческий интеллект оперирует не объектами, а внутренними образами, то есть отражениями объектов в системе. Внутренние образы описываются в пространстве потребностей, в самом примитивном случае — одномерном, а в случае реальных живых организмов, либо в случае реализации для систем ИИ со сложной иерархией взаимозависимых потребностей — тензорном, именуется «матрицей потребностей» согласно предложенной ранее архитектуре [22].

Для решения основной задачи – выживания индивидуального экземпляра, социума или вида - целью системы становится удовлетворение текущих потребностей в соответствии с выбранным приоритетом. Динамика функциональных потребностей может описываться математикой рыночной экономики, включая управление рисками через вероятностный анализ исходов альтернативных сценариев [19]. Взаимозависимость между физиологическими и психологическими потребностями рассчитывается через психологическую ценность носителя интеллекта как потенциальную энергию выживания - удовлетворения базовых потребностей/мотиваций, где выживание только одна из них.

Поток потребностей непрерывен. Согласно Маслоу [25], удовлетворение одной потребности ведет к появлению новой, потребности конкурентны и взаимозависимы. Также на конкурентность и взаимозависимость потребностей указывал П.К. Анохин [8]. В свою очередь, цель отличается от потребности тем, что она конечна. Цель заканчивается, когда она достигнута.

Значение приоритетов текущих потребностей определяет текущее состояние системы. Текущие значения удовлетворенности, неудовлетворенности и востребованности (приоритета) потребности, характеризуют текущую точку (координату) в функциональном пространстве целенаправленной человеческой деятельности в перспективе покрытия Маркова [23].

Высшие потребности можно выразить через комбинацию базовых потребностей [25]. Поэтому любую цель системы можно описать как комбинацию базовых потребностей и потребностей младших порядков. Пространство («матрица») потребностей (ПП) – основной элемент системы, на котором строятся все расчёты и сравнения. Потребности отражают текущее состояние системы. По сути ПП это набор строк строк по уровням состояния системы (как в программе-супервизоре операционной системы), где в каждой ячейке содержится текущее значение приоритета потребности. ПП – универсальный математический инструмент для описания деятельности модели человека при описании его поведения либо системы ИИ при ее реализации.

В случае моделирования человеческой психики, ПП отражает смысл любого объекта для человека в природе состоит в возможностях удовлетворения его потребностей в комбинации разных уровней. В этом случае, ПП может может быть декомпозировано на уровни: а) Индивидуальный (Базовый); б) Семейно-бытовой; в) Социально-экономический; г) государственно-цивилизационный. Каждый уровень имеет свой приоритет при принятии решений, причем приоритеты могут быть различными для различных индивидуумов.

Представляется возможным создание математической модели, позволяющей осмыслить картину мира в контексте ПП и найти оптимальное решение для деятельности модели живого интеллектуального организма либо экземпляра искусственной интеллектуальной системы. Алгоритм анализа может быть построен на выборе самого эффективного решения текущей задачи с учетом рисков, ограниченности ресурсов и в рамках горизонта планирования.

### Математическая модель

Если рассматривать функционирование человеческой психики в течение всего жизненного цикла, а также искусственной психики гипотетического интеллектуального агента как обучение на основе получаемого опыта взаимодействия с окружающей средой, согласно [26], то в разработках систем искусственного интеллекта; наиболее распространенной постановкой является обучение с подкреплением [27]. Оно основывается на Марковской модели процесса принятия решений [28], причем явно получаемое агентом подкрепление в ходе реализуемого им поведения используется для определения так называемой «полезности» того или иного решения или действия, предшествовавшего подкреплению, в свете «теории ожидаемой» полезности [29]. Однако в экономике, на сегодняшний день, последняя теория уступила место «теории перспектив» [19], согласно которой, полезность действия оценивается как позитивными, так и негативными исходами от его совершения, так и различными вероятностями, ассоциированными с этими исходами, причем субъективная полезность позитивных и негативных исходов может быть различна и эти различия могут быть отличны у различных субъектов. На практике, при обучения на основе опыта, приходится также оценивать каждое решение или действие с точки зрения не просто отдельного подкрепления, а комбинации гипотетических явных подкреплений - за достижение различных целей (получение пищи или удовольствия) или же неявных - за избегание угроз (удара током или обливания водой), или же получение полезного жизненного опыта либо достижение определенного социального статуса, что может потребоваться для достижения целей и избегания угроз в отдаленном будущем, но не может быть «подкреплено» в ближайшей перспективе. Таким образом, требуется расширение скалярной оценки оценки полезности с заменой ее на векторную, как например, в [26] полезность рассчитывается в двухмерном пространстве положительного и отрицательного подкрепления — как полезности получения выгод и избегания потерь согласно «теории перспектив» [19]. Введение необходимости явной поддержки исследовательской активности может быть введением информационной полезности за обнаружение новых паттернов, а повышение эффективности потребления энергии — за счет энергетической полезности. В сложных системах, таких как человек или высоко-интеллектуальный агент, пространство потребностей может быть структурировано в гетерархию взаимосвязанных потребностей и описываться не вектором, а матрицей, где строки матрицы соответствуют потребностям различных уровней от базовых до высших [25] или, в общем случае, тензором. Тем самым, может быть формализован принцип интеллекта как средства достижения сложных целей по Герцелю [3] и Вонгу [4] — определением сложной тензорной функции полезности.

Понятие сложной цели, в контексте Рис.1, может быть формализовано как «мотивационный вектор» $z$ [30]. На основе определенного нами пространства потребностей, этот вектор может быть определен как скалярное произведение двух векторов - а) долгосрочной жизненной важности (приоритета) потребности $x$ и б) краткосрочной неудовлетворенности (актуализации) данной потребности $y$. Вектор приоритизации $x$ может рассматриваться как генетический и культурный код агента, предопределенный его генетикой (в случае человека) либо встроенный (в случае искусственного ИА), а также приобретенный в ходе его развития и долгосрочного обучения («формирования личности»). Он определяет, например соотношение склонностей к рискам ради приобретений (положительных подкреплений) и избегания потерь

(отрицательных подкреплений) в свете теории перспектив [19], стремлению к обеспечению энергоэффективности [8, 16, 17, 20, 21], и потребности к познанию, удовлетворяемой как повышение предсказуемости окружающей реальности, согласно [15]. В свою очередь, вектор актуализации определяется на основе данных, получаемых в виде эмоций [2, 6] как неудовлетворенность соответствующих потребностей в конкретный момент принятия решений.

Таким образом, в рамках Марковской модели процесса принятия решений [28], ожидаемая ценность или полезность реализации того или иного вектора действий и бездействий $a$ в текущем состоянии $s$, для момента времени $t$, а также подкрепления, получаемого при обучении [26, 27] в очередной момент времени $t+1$, может определяться следующими векторными и скалярными функциями.

$s_t$ — вектор переменных состояния, со значениями определяемыми, согласно Рис.1, посредством сенсоров агента, в том числе — включающих: сенсоры актуализации потребностей $y$, например — воспринимающие положительные и отрицательные подкрепления, приобретение либо потерю энергетического ресурса, а также подтверждение ожидания результатов предсказания или действия (акцепторы результата действия [8]); сенсоров фиксации совершения действия $a$; сенсоров ощущений $f$, не связанных явно с потребностями или деятельностью. Соразмерный вектор $s'_{t+1}$ определяет ожидания.

$a_t$ — вектор действий, совершенных агентом, либо планируемых к совершению в очередной момент времени $a'_{t+1}$, причем совершенность действий также фиксируется соответствующими переменными состояния ($a \subseteq s$).

$y_t$ — вектор актуализации потребностей, выражаемых через «выделенные» переменные состояния, причем актуализация потребностей является частью состояния ($y \subseteq s$), либо сформированных ожиданий $y'_{t+1}$.

$f_t$ — вектор ощущений, не связанных непосредственно с актуализацией и удовлетворением потребностей либо совершением действий ($f \subseteq s$).

$x$ — вектор приоритизации потребностей, предполагаемый неизменным для агента в краткосрочной перспективе, но может быть изменен в долгосрочной.

$z_t = x \cdot y_t$ — вектор мотивации к удовлетворению потребностей с учетом их приоритизации и актуализации («мотивационный вектор» [30]).

$U(s,s')$ — скалярная функция (матрица) полезности реализации состояния $s'$ (включая совершаемые при этом действия $a'$ либо без совершения каких-либо действий вообще) при исходном состоянии $s$ в контексте приоритизации потребностей *x* агента, определяемая в ходе обучения на основе подкреплений $r_{t+1}$ на переходах из состояния $s_t$ в $s_{t+1}$, соответствует Q-функции в случае применения метода Q-learning [31] (в последнем случае функция полезности вырождается в *U(a,a')*).

$r_{t+1} = x \cdot (y_t - y_{t+1})$ — скалярная функция подкрепления (по существу - эмоции) в результате действия или бездействия $a_{t+1}$ при переходе из состояния $s_t$ в состояние $s_{t+1}$, включая «выделенные» потребности: 1) энергоэффективности *e(a)*; 2) предсказуемости как функции от разности $s'_{t+1} - s_{t+1}$ («план минус факт») между ожиданиями $s'_{t+1}$, сформированными в момент времени *t*, и фактическим состоянием $s_{t+1}$ в момент времени *t+1*.

*Q(s,s')* и *P(s,s')* — матрицы взаимоисключения (*Q*) либо взаимообусловленности (*P*) указанных выше переменных *s*, если какие то из них физически не могут быть удовлетворены (*y*), реализованы (*a*) или испытаны (*f*) агентом одномоментно (*Q*) или же — какие должны быть удовлетворены, совершены либо испытаны одновременно или одно необходимо для другого (*P*).

## Минимальная экспериментальная реализация

В качестве эксперимента была проведена минимальная реализация обучения агента игры в теннис от стенки («Self Pong») с расширением модели, описанной и реализованной в [26], открытый код опубликован в https://github.com/aigents/aigents-java/tree/master/src/main/java/net/webstructor/agi

В реализованном эксперименте поведение агента осуществлялось в 4-мерном пространстве потребностей: 1) положительного подкрепления при отражении мячика либо удара его об стенку (Happy); 2) избегания отрицательного подкрепления при ударе мячика об стенку на стороне игрока с ракеткой; 3) новизны обнаруживаемых агентом состояний (Novelty); ожидаемости или предсказуемости переживаемых ситуаций (Expectedness). При этом измерялась и визуализировалась актуализация всех четырех потребностей (Рис 2.), и исследовались различные значения векторов приоритизации для первых двух из них, связанных с подкреплением. Эксперимент проводился с использованием различных стратегий обучения на основе опыта, включающего как явное подкрепление по нескольким каналам извне, так и удовлетворение внутренних потребностей агента, описанных ранее [26]. При этом стратегии обучения предполагали формирование эпизодической памяти либо в виде сегментов последовательностей $s_t$ и $a_t$ между подкреплениями (положительными и отрицательными), либо в виде карт переходов между состояниями $s_t$ и $a_t$, а подкрепление $R_t$ изменяло значения полезности всех элементов эпизодической памяти на сегментах между подкреплениями («глобальное подкрепление») [26], в отличие от Q-learning [31], где подкрепление изменяется инкрементально, распространяясь по оси времени от будущих состояний к предыдущим.

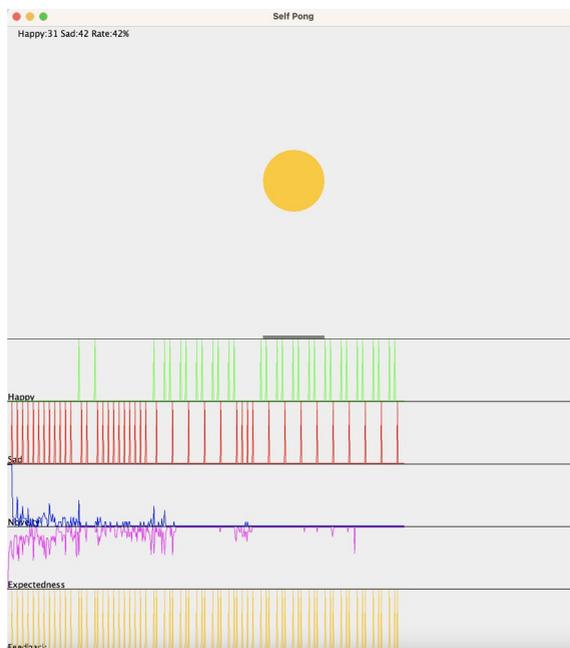

Рис. 2. Визуализация процесса обучения агента игре в теннис от стенки в пространстве потребностей - стенка наверху, ракетка в середине, мячик между стенкой и ракеткой, внизу функции удовлетворенности либо неудовлетворенности по четырем потребностям (Happy, Sad, Novelty, Expectedness) и отдельная функция явного (положительного либо отрицательного) подкрепления (Feedback)

В результате эксперимента, для данной задачи был подтвержден эффект негативного влияния отрицательного подкрепления на обучаемость агента. То есть, при равенстве приоритетов потребностей получения положительного и избегания отрицательного подкрепления и соответствующего формирования значений функций ожидаемой полезности в модели, обучение замедлялось и для некоторых моделей и конфигураций игрового поля и отложенности подкрепления становилось невозможным ввиду подавления исследовательской активности агента наказаниями за сделанные ошибки при проверке новых стратегий.

## Заключение

Интеллект человека всегда отвечает на вопрос: "Лучше или хуже станет от принятия решения". Экономика удовлетворяет потребности человека - физические и психологические в его социально-производственной деятельности. Психика управляет удовлетворением потребностей на уровне поведения человека как индивидуума.

С помощью соответствующей математической модели мы можем считать степень удовлетворения потребностей в системе через затраты энергии для деятельности и времени - как на уровне экономики, так и на уровне психики. Что и является основной задачей интеллекта.

Основа решения задачи выживания - непрерывное обеспечение жизненной энергией. Деятельность человека требует затрат энергии. Таким образом, получается взаимозависимость получения и эффективного использования энергии. Интеллект позволяет выбрать оптимальное решение. Чем выше интеллект, тем эффективнее результат деятельности. Значение коэффициента интеллекта состоит в том, сколько интеллектуального труда было затрачено ранее для создания и получения данного объекта или знания.

Для обеспечения выживания вида, социума или индивида человек осуществляет целенаправленную деятельность, опираясь на имеющиеся знания и опыт, приобретенный индивидуально либо коллективно, максимизируя значение коэффициента интеллекта как индивидуума, так и социума. В свою очередь, при создании и сравнении искусственных интеллектуальных систем, максимизация указанного коэффициента может рассматриваться как основной показатель эффективности системы.

Способность интеллектуального агента осуществлять свою деятельность в сложных средах с учетом множества целей и угроз, определяемых в контексте индивидуального пространства биологических, физических и экзистенциальных потребностей, с оптимизацией удовлетворения последних, может быть рассмотрено как общий интеллект, сила которого определяется размерностью и сложностью этого пространства.

Вычислительная реализация когнитивной архитектуры на основе предлагаемого формализма позволяет получить предварительные результаты, убедительные с точки зрения их интерпретации.

*Список литературы*

1. Норберт Винер. "Человек управляющий". СПб.: Питер, 2001. – 288 с. – (Серия «Психология-классика»). ISBN 5-318-00214-5.
2. Леонтьев Д.А. Л478 Психология смысла: природа, строение и динамика смысловой реальности. 2-е, испр. изд. — М.: Смысл, 2003. — 487 с.
3. Goertzel, B. "The general theory of general intelligence: A pragmatic patternist perspective". Preprint at https://arxiv.org/abs/2103.15100, 2021.
4. Wang, P. "Artificial intelligence: What it is, and what it should be". C. Lebiere & R. Wray (Eds.), Papers from the AAAI spring symposium on between a rock and a hard place: Cognitive science principles meet ai-hard problems, pp. 97–102. AAAI Press, 2006.
5. Даниэль Канеман. "Думай медленно… решай быстро" АСТ, 2023.
6. Павел Васильевич Симонов. "Эмоциональный мозг" © ООО Издательство «Питер», 2021.
7. Дубынин В.А. "Мозг и его потребности. 2.0. От питания до признания". Альпина нон-фикшн, 2024.
8. Анохин, П.К. Принципиальные вопросы общей теории функциональных систем / Москва : Директ-Медиа, 2008. 131 с. ISBN 978-5-9989-0384-7.
9. Л. фон Берталанфи. Общая теория систем: критический обзор. В сборнике переводов Исследования по общей теории систем. М.: Прогресс, 1969. 520 с.
10. Vityaev, E., & Demin, A. "Cognitive architecture based on the functional systems theory". Procedia Computer Science , 145 , 623-628, https://doi.org/10.1016/j.procs.2018.11.072 (Postproceedings of the 9th Annual International Conference on Biologically Inspired Cognitive Architectures, BICA 2018 (Ninth Annual Meeting of the BICA Society), Prague, Czech Republic), 2018.
11. Anton Kolonin, Evgenii Vityaev, Yuriy Orlov. "Cognitive Architecture of Collective Intelligence Based on Social Evidence", Procedia Computer Science, Volume 88, 2016, Pages 475-481, ISSN 1877-0509, https://doi.org/10.1016/j.procs.2016.07.467.
12. Cisek, P. "Cortical mechanisms of action selection: the affordance competition hypothesis". Philosophical Transactions of the Royal Society B: Biolog- ical Sciences, 362(1471), 1585–1599, https://doi.org/10.1098/rstb.2007.2054 Retrieved from http://doi.org/10.1098/rstb.2007.2054, 2007.
13. Wang, P. "Non-axiomatic reasoning system: exploring the essence of intelligence" (Unpublished doctoral dissertation). Indiana University, USA. (UMI Order No. GAX96-14571), 1996.
14. Kolonin, A. (2015). Computable cognitive model based on social evidence and restricted by resources: Applications for personalized search and social media in multi-agent environments. International Conference on Biomedical Engineering and Computational Technologies (SIBIRCON), 2015, Novosibirsk, Russia, https://ieeexplore.ieee.org/document/7361869.
15. Friston, K. "The free-energy principle: a unified brain theory?" Nature Reviews Neuroscience, 11(2), 127-138, https://doi.org/10.1038/nrn2787, 2010, Feb 01.

16. Goertzel, I.F., Heljakka, A. "Probabilistic logic networks: A comprehensive framework for uncertain inference". New York, NY: Springer, 2008.
17. Vityaev, E.E., Perlovsky, L.I., Kovalerchuk, B.Y., Speransky, S.O. "Probabilistic dynamic logic of cognition". Biologically Inspired Cognitive Architectures, 6, 159-168, https://doi.org/10.1016/j.bica.2013.06.006 (BICA 2013: Papers from the Fourth Annual Meeting of the BICA Society), 2013.
18. Dawid, A., & LeCun, Y. "Introduction to latent variable energy-based models: a path toward autonomous machine intelligence". Journal of Statisti- cal Mechanics: Theory and Experiment, 2024(10), 104011, https://doi.org/10.1088/1742-5468/ad292b, 2024, October.
19. Daniel Kahneman and Amos Tversky "Prospect Theory: An Analysis of Decision under Risk, Econometrica Vol. 47, No. 2 (Mar., 1979), pp. 263-292 (29 pages) Published By: The Econometric Society
20. Владимир Михайлович Бехтерев. "Объективная психология". Выпуск 1. 1907.
21. Зигмунд Фрейд. "По ту сторону принципа удовольствия". АСТ, 2021.
22. Владимир Германович Крюков. САМООБУЧЕНИЕ СИСТЕМЫ ПРИНЯТИЯ РЕШЕНИЙ В МУЛЬТИАГЕНТНОЙ СРЕДЕ. Патент RU 2830819 C1. Дата подачи заявки: 09.06.2023, Опубликовано: 26.11.2024 Бюл. № 33
23. Gorban, Alexander N.; Gorban, Pavel A.; Judge, George. "Entropy: The Markov Ordering Approach" (http://www.mdpi.com/1099-4300/12/5/1145/). Entropy (http://www.mdpi.com/journal/entropy) 12, no. 5 (2010), 1145—1193.
24. Альфред Адлер. "Понять природу человека". Alfred Adler MENSCHENKENNTNIS © Перевод. Е. Цыпин, 2020 © Издание на русском языке AST Publishers, 2021
25. Абрахам Маслоу. "Новые рубежи человеческой природы". Альпина нон-фикшн, 2011.
26. Kolonin, A. (2022). Neuro-Symbolic Architecture for Experiential Learning in Discrete and Functional Environments. In: Goertzel, B., Iklé, M., Potapov, A. (eds) Artificial General Intelligence. AGI 2021. Lecture Notes in Computer Science(), vol 13154. Springer, Cham. https://doi.org/10.1007/978-3-030-93758-4_12
27. van Otterlo, M., Wiering, M. (2012). Reinforcement Learning and Markov Decision Processes. In: Wiering, M., van Otterlo, M. (eds) Reinforcement Learning. Adaptation, Learning, and Optimization, vol 12. Springer, Berlin, Heidelberg. https://doi.org/10.1007/978-3-642-27645-3_1
28. Puterman, Martin L. (1994). Markov decision processes: discrete stochastic dynamic programming. Wiley series in probability and mathematical statistics. New York: Wiley. ISBN 978-0-471-61977-2
29. John von Neumann and Oskar Morgenstern. (1944). Theory of Games and Economic Behavior. Princeton University Press.
30. V. F. Petrenko and A. P. Suprun, "Goal oriented systems, evolution, and the subjective aspect in systemology," Tr. Inst. Sistem. Analiza RAN 62 (1) (2012)
31. Li, Shengbo (2023). Reinforcement Learning for Sequential Decision and Optimal Control (First ed.). Springer Verlag, Singapore. pp. 1–460. doi:10.1007/978-981-19-7784-8. ISBN 978-9-811-97783-1. S2CID 257928563